\documentclass[12pt]{article}
\usepackage{graphicx}
\usepackage{amsfonts}
\usepackage{pstricks}
\def\beq{\begin{equation}}   \def\eeq{\end{equation}}

\begin{document}
\begin{titlepage}
\begin{flushright}
TPI-MINN-00/27-T  \\
UMN-TH-1908/00   \\
FTUV/000601 \\
IFIC/00-32 \\
\end{flushright}

\vspace{0.6cm}

\begin{center}
\Large{{\bf Leaving the BPS bound: Tunneling of classically 
saturated solitons}}
\end{center}

\vspace{1cm}

\begin{center}
\large D. Binosi
\end{center}

\begin{center}
{\em
Departamento de F\'\i sica Te\'orica and IFIC, Centro Mixto, \\
Universidad de Valencia-CSIC,\\
E-46100, Burjassot, Valencia, Spain\\}
\end{center}

\vspace{0.1cm}

\begin{center}
\large M. Shifman and T. ter Veldhuis
\end{center}

\begin{center}
{\em Theoretical Physics Institute, Univ. of Minnesota, 
Minneapolis,
       MN 55455}
\end{center}
\vspace{0.5cm}

\begin{abstract}
We discuss quantum tunneling between classically BPS saturated solitons 
in  two-dimensional theories with ${\cal N}=2$ supersymmetry and
a compact space dimension. Genuine BPS states form  shortened
multiplets of dimension two. In the models we consider there are two
degenerate  shortened multiplets at the classical level, but  
there is no obstruction to pairing up through quantum tunneling. 
The tunneling amplitude in the imaginary
time is described by instantons. We find that the
instanton is nothing but the $1/4$ BPS saturated ``wall junction,'' considered
previously in the  literature in other contexts. Two central charges of the
superalgebra allow us to calculate the instanton action without finding the
explicit solution (it is checked, though, numerically, that the saturated
solution does exist). We present a quantum-mechanical interpretation of the
soliton tunneling.

\end{abstract}

\vspace{0.5cm}

\begin{flushleft}
E-mail: binosi@titan.ific.uv.es, shifman@physics.spa.umn.edu, 
veldhuis@hep.umn.edu
\end{flushleft}
\end{titlepage}

\section{Introduction}

Bogomol'nyi-Prasad-Sommerfield (BPS) saturated topologically stable solitons in
supersymmetric theories are widely discussed at present in  connection with the
brane world scenarios \cite{DS,ADD}. In theories with compact spatial
dimension(s), two distinct degenerate mass solitons which are BPS saturated
classically, and to any finite order in  perturbation theory, can mix
nonperturbatively, thus lifting the BPS bound. Two shortened supermultiplets
pair up with each other combining in one full supermultiplet with mass $M>|Z|$,
where $Z$ is the central charge of the superalgebra. This phenomenon is an
analog (in the soliton sector) of the spontaneous breaking of  supersymmetry
due to instantons in the vacuum sector \cite{MV,ADS}. To the best of our
knowledge it was first considered in the context of ${\cal N}=2$
two-dimensional Wess-Zumino models in \cite{HLS}.

In this work we address the issue of calculating the shift $M-|Z|$. In the
quasi-classical approximation, it is proportional to the tunneling probability
which, in turn, is determined by instantons. Remarkably, the instanton calculus
in this case is nothing but an adaptation of the theory of the BPS saturated
wall junctions, which also received much attention recently
\cite{AGIT,AT,GT,CHT,GS,BV,SV}. In particular, the instanton action can be
derived from the central charges. The explicit formula for the instanton
solution is not needed. The only thing we need to know is the very fact of its
existence. This is in perfect parallel with the standard (non-supersymmetric)
instantons: once one knows that the self-duality equations have a solution, the
instanton action is unambiguously fixed in terms of the topological charge.

In Ref.~\cite{HLS} it was observed that the soliton mixing, resulting in the
loss of the BPS saturation, can be described by an effective SUSY quantum
mechanics; however, the general construction presented there, is not very
transparent. Here we reduce the construction of Ref.~\cite{HLS} to a simple
limiting case which nicely illustrates the essence of the phenomenon.

The organization of the paper is as follows. In Section~2 we formulate the
problem and elaborate general aspects of the solutions. In Section~3 a specific
instructive example is considered. Section~4 is devoted to SUSY quantum
mechanics.

\section{Formulation of the problem and general results}
\label{sec:generalresults} 

Classically BPS saturated soliton supermultiplets which may become degenerate
in mass with some other supermultiplets and lift the BPS bound because of a
nonperturbative mixing, is a rather general feature of  various theories.
Although our results are applicable in all cases, we find it convenient to
explain the problem in a specific setting.

Consider a two-dimensional ${\cal N}=2$ Wess-Zumino model of one chiral
superfield $\Phi$ with the superpotential ${\cal W}(\Phi)$. Any model of this
type can be obtained as a two-dimensional reduction of the corresponding 
four-dimensional theory. The geometry of the world sheet is a cylinder. As
explained in \cite{HLS}, for the existence of the BPS solitons it is necessary
that ${\cal W}(\Phi)$ is a multi-branch function (otherwise the vanishing of
the central charge $Z$ cannot be avoided), while $d{\cal W}/d\Phi$ must be
meromorphic. Another necessary (and sufficient) condition for the topologically
stable solitons is the existence of non contractable cycles in the target
space. One of the simplest choices is a target space with the topology of a
cylinder, possibly with punctured points. Then the periods of $\cal W$ are the
central charges  of the SUSY algebra,
\begin{equation}
Z_i =  2 \oint_{{\rm nc}_i} d{\cal W}, \label{cs}
\end{equation}
where nc$_i$ stands for the $i$-th non-contractable contour in the target space.

\setlength{\unitlength}{.1cm}
\begin{figure}[!t]
\begin{center}
\includegraphics[width=14.0cm]{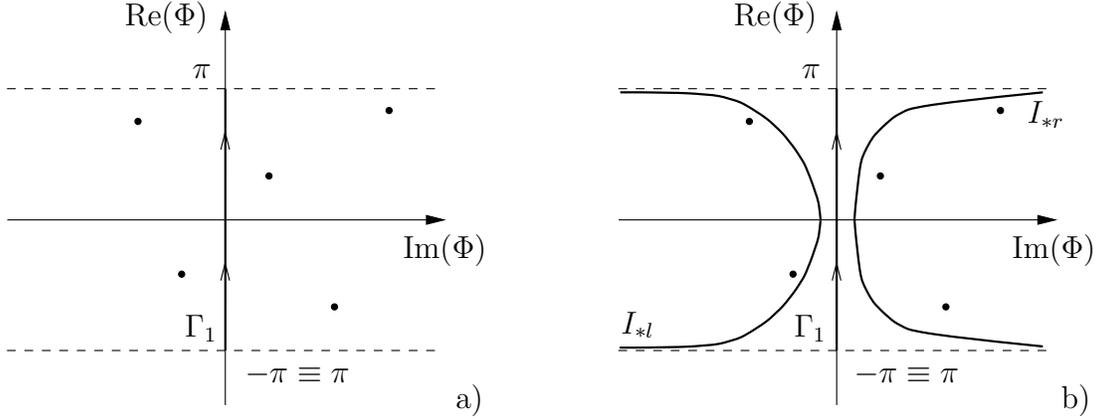}
\put(-87,20){Im$(\Phi)$}
\put(-6,20){Im$(\Phi)$}
\put(-124,51){Re$(\Phi)$}
\put(-43,51){Re$(\Phi)$}
\put(-108,3.5){$-\pi\equiv \pi$}
\put(-27,3.5){$-\pi\equiv \pi$}
\put(-115,44){$\pi$}
\put(-34,44){$\pi$}
\put(-116,9.5){$\Gamma_1$}
\put(-35,9.5){$\Gamma_1$}
\put(-58,10){$I_{*l}$}
\put(-4,38){$I_{*r}$}
\put(-80,0){a)}
\put(0.5,0){b)}
\caption{a) The general singularity structure of the quantity
$d{\mathcal W}/d\Phi$ in the complex $\Phi$ plane; b) the critical
trajectories corresponding to infinite periods.}
\label{fig1I}
\end{center}
\end{figure}

If there is at least one non-contractable cycle, one can always define
$\Phi$ in such a way that $d {\cal W}$ is periodic
\begin{equation}
d {\cal W} (\Phi+2 \pi) = d {\cal W} (\Phi).
\end{equation}
This particular parametrization is not crucial, and is imposed only for the
purpose of making the presentation simpler. The poles of $d {\cal W}/d\Phi$ are
assumed to be single poles; quadratic and higher order poles can be treated as
a limiting case of  coinciding single poles. A generic singularity structure of
$d {\cal W}/ d\Phi$ is depicted in Fig.~\ref{fig1I}a), where the poles are
marked by bold dots. The K\"{a}hler potential is taken to be trivial, ${\cal
K}(\Phi,\bar{\Phi}) = \Phi \bar{\Phi}$. 

\begin{figure}[!t]
\begin{center}
\includegraphics[width=6.5cm]{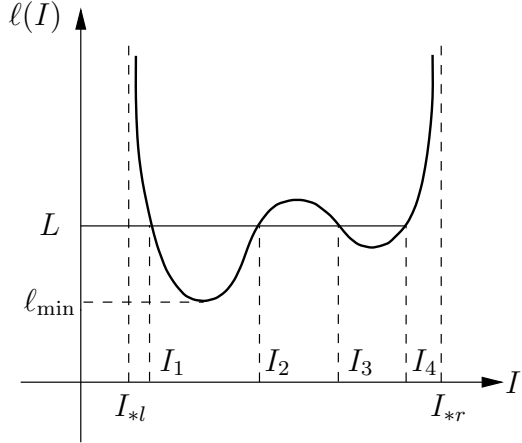}
\put(-66,56){$\ell(I)$}
\put(-62,28){$L$}
\put(-64,18){$\ell_{\mathrm{min}}$}
\put(-52,4){$I_{*l}$}
\put(-10,4){$I_{*r}$}
\put(0,7){$I$}
\put(-46,10){$I_1$}
\put(-32,10){$I_2$}
\put(-21,10){$I_3$}
\put(-12.5,10){$I_4$}
\label{fig2I}
\caption{The period function $\ell$ versus $I$.}
\end{center}
\end{figure}

Consider the cycle $\Gamma_1$, for which 
\begin{equation}
Z_1 = 2 \int_{\Gamma_1}\!d\Phi\left(\frac{d {\cal W}}{d\Phi}\right).
\end{equation}
If $Z_1 = |Z_1| e^{\imath \delta}$, then the equation for the static
BPS soliton has the form
\begin{equation}
\frac{d\Phi}{dx} = e ^{\imath \delta } \frac{d\bar{\cal W}}{d\bar{\Phi}}. 
\label{BPSeq}
\end{equation}
This equation admits the \lq\lq integral of motion\rq\rq,
\begin{equation}
I  =  {\rm Im} \left(e^{-\imath \delta} {\cal W} \right), 
\end{equation}
{\it i.e.} $dI/dx=0$ when $\cal W$ and $\bar{\cal W}$ are evaluated on the
solution of (\ref{BPSeq}). The existence of this integral of motion allows one
to find the BPS solution in the general case~\cite{HLS}. The strategy is as
follows. We first ignore that the world sheet is a cylinder with period $L$ in
the $\hat{x}$ direction, and solve the BPS equation without posing the
condition of periodicity, $\Phi(x+L)=\Phi(x)$. The solution found in this way
is marked by the continuous (real) parameter $I$, and is a periodic function of
$x$ with period
\begin{equation}
\ell(I)=\int d\Phi \left(e^{\imath \delta} \frac{d\bar{\cal W}}{d\bar{\Phi}}
\right)_{\bar{\Phi}=\bar{\Phi}_I(x)}^{-1}.
\end{equation}
The period function $\ell(I)$ is real and positive. It is not difficult to show
that there exist critical values of $I$ such that $\ell(I)\rightarrow\infty$ at
$I\downarrow I_{*l}$ or $I\uparrow I_{*r}$, where $I_{*l,r}$ mark the critical
trajectories running close to the nearest poles of $\cal W$ from the left and
from the right (Fig.~\ref{fig1I}b). A schematic plot of the function $\ell(I)$
is presented in Fig.~\ref{fig2I}.

If the circumference of the worldsheet cylinder $L>\ell_{\mathrm{min}}$,  then
the equation $\ell(I)=L$ has an even number of solutions. The corresponding
values $I_i$ belongs to the interval  $(I_{*l}, I_{*r})$, while $\Phi_{I_i}(x)$,
$i=1,2,...,2\nu$ are the classical  BPS solutions satisfying Eq.~(\ref{BPSeq})
and the periodicity  condition $\Phi_{I_i}(x+L)=\Phi_{I_i}(x)$.

In the case at hand they have particle interpretation. Altogether, we have
$2\nu$ supermultiplets, each containing two states. All masses are degenerate
and equal to $|Z_1|$.

The BPS nature of the solitons established above at the classical level
persists to any finite order in perturbation theory (this statement assumes
that there is a small expansion parameter in the superpotential and/or the
K\"{a}hler function). Alternatively, one can say that $2\nu$ BPS solitons
remain under small deformations of the parameters. This is due to the fact that
the number of states in the supermultiplet is two, while the full ${\cal N}=2$
supermultiplet contains four states. The BPS supermultiplets are shortened.

It is equally clear, however, that nonperturbatively the BPS saturation of the
solitons under consideration is lifted, they pair up to form $\nu$ full
supermultiplets which lift the BPS bound. This was noted in~\cite{HLS}, where
arguments were given based on the Cecotti-Vafa-Intriligator-Fendly (CVIF) 
index~\cite{CVIF}. Our task here is to calculate $M-|Z_1|$ in the
quasi-classical approximation (which implies of course that $(M-|Z_1|)/|Z_1|
\ll 1$).

The BPS saturation is lifted by tunneling. Consider for simplicity the case
when Eq.~(\ref{BPSeq}) has only two solutions, $\Phi_{I_1}(x)\equiv\phi_1(x)$
and $\Phi_{I_2}(x)\equiv\phi_2(x)$. One can construct an interpolating field 
configuration $\phi(t,x)$ (where $t$ is the Euclidean time) such that in the
distant past
\begin{equation}
\phi(t,x)\,{\buildrel t=-T/2\to-\infty\over{\hbox to
60pt{\rightarrowfill}}}\,\phi_1(x), 
\label{bc1}
\end{equation}
and in the distant future
\begin{equation}
\phi(t,x)\,{\buildrel t=T/2\to\infty\over{\hbox to
60pt{\rightarrowfill}}}\,\phi_2(x),
\label{bc2}
\end{equation}
The interpolation is smooth (in particular,  $\phi(t, x+L)=\phi(t,x)$ for all
$t$), so that the 
(Euclidean) action $A$
\begin{equation}
A\left[\phi(t,x)\right]- |Z| T
\end{equation}
is finite. Here $|Z|$ is the soliton mass in the absence of the tunneling. The
quasi-classical formalism is applicable provided $A-|Z|T \gg 1$. One must
minimize over all interpolating trajectories; once the trajectory $\phi_0(t,x)$
corresponding to the minimal action is found, one
can calculate
\begin{equation}
\Delta A_{\rm min} = A\left[ \phi_0(t,x)\right] - |Z| T.
\end{equation}
The shift of the soliton masses from the BPS bound is then
\begin{equation}
\Delta M = M - |Z| \propto e^{-2 \Delta A_{\rm min}},
\end{equation}
where the factor of $2$ in the exponent is due to the fermion zero modes. We
will comment more on this factor in Section~4.

The central result of the present work is as follows. The determination of the
minimizing trajectory $\phi_0(t,x)$  (in the Euclidean time) reduces to the
problem of determining the BPS wall junction in the  $(1+2)$-dimensional theory,
in which the  $(1+1)$-dimensional model under consideration is embedded. The
embedding is trivial. Indeed, if the original model is a $(1+1)$-dimensional
slice of the four dimensional Wess-Zumino model, the one in which we embed is a
$(1+2)$-dimensional slice of the very same model. From this remark it is clear
that the formalism we discuss is applicable, generally speaking, in the
supersymmetric theories with extended supersymmetry (${\cal N}=2$ or ${\cal
N}=4$). In fact, since the solitons -- the ``walls" and ``wall junctions" --
are static, we will have to deal with one- and two-dimensional problems,
respectively.

\begin{figure}[!t]
\begin{center}
\includegraphics[width=6.0cm]{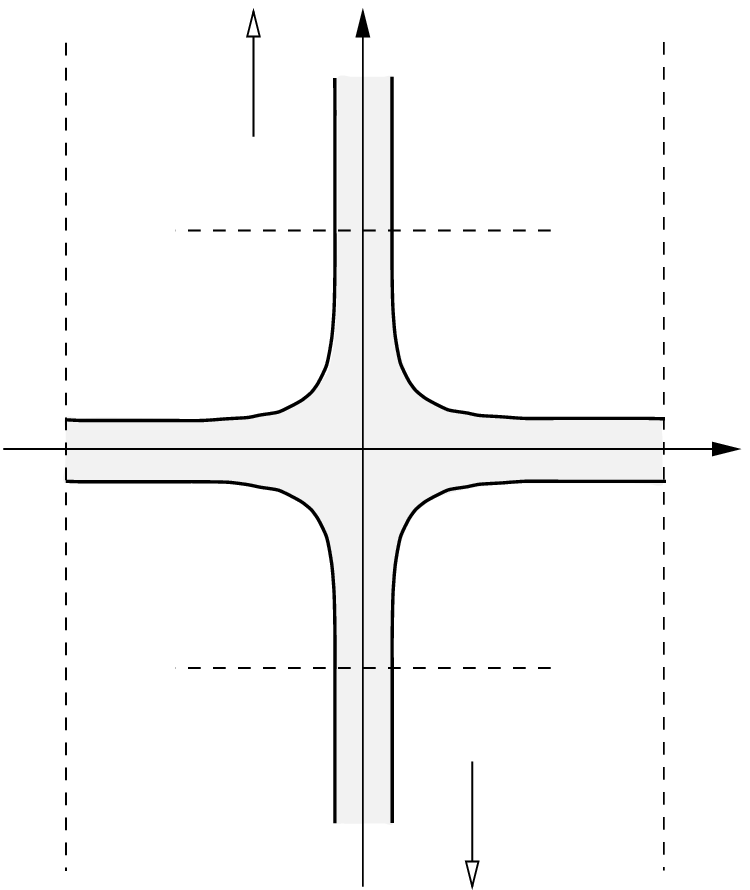}
\put(-2,36.5){$x$}
\put(-34,69.5){$t$}
\put(-5,31){$L/2$}
\put(-67,37.5){$-L/2$}
\put(-20,55){$\phi_2(x)$}
\put(-48.5,20){$\phi_1(x)$}
\put(-20,4){$-T/2$}
\put(-48.5,64){$T/2$}
\label{fig3I}
\caption{The energy distribution corresponding to the interpolating
configuration; the unshaded area has the \lq\lq vacuum\rq\rq\ energy
density.}
\end{center}
\end{figure}
 
The energy distribution for the interpolating configuration $\phi_0(t,x)$ is
schematically depicted in Fig.~\ref{fig3I}. It is nothing but an adaptation of
the standard four-wall junction on the cylinder world sheet. The equation for
the BPS wall junction has the form
\begin{equation}
\frac{\partial \phi}{\partial \zeta} = \frac{1}{2} \frac{d \bar{\mathcal W}}{d
\bar{\phi}}, 
\label{BPSjunc}
\end{equation}
where $\zeta$ is the complex variable $\zeta=x+it$, and 
$\partial_\zeta=1/2(\partial_x-\imath\partial_t)$. The solution
to this equation, if it exists, is $1/4$ BPS saturated.
Equation~(\ref{BPSjunc}) was
first derived in~\cite{CS}. The fact that the solution of Eq.~(\ref{BPSjunc})
minimizes the Euclidean action  is quite obvious. Indeed,
\begin{eqnarray}
A & = & \int dt\,dx\, \left(\left|\frac{\partial \phi}{\partial t}\right|^2
+\left|\frac{\partial \phi}{\partial x}\right|^2 + \left|\frac{d {\cal W}}{d \phi}\right|^2
\right), \\
 & = & \int dt\,dx\, \, \left( 2 \frac{\partial \phi}{\partial \zeta} -
\frac{d\bar{\cal W}}{d\bar{\phi}} \right) \left(2 \frac{\partial \bar{\phi}}
{\partial \bar{\zeta}} - \frac{d {\cal W}}{d \phi} \right)+
\,\, {\rm surface}\,\,\, {\rm terms}. \nonumber \label{euclact}
\end{eqnarray}
The surface terms are unambiguously fixed by the boundary conditions,
Eqs.~(\ref{bc1}) and~(\ref{bc2}), and by the periodicity condition 
$\phi(t,x+L)=\phi(t,x)$. Thus, the construction is quite analogous to the
instanton self-duality equation in the Yang-Mills theory or the two-dimensional
$O(3)$ sigma model, where the surface terms are topological charges.

In the supersymmetric model under consideration the surface terms are 
proportional to two distinct central charges which exist in the
superalgebra~\cite{GS}.  One central charge is related to the junction \lq\lq
spokes\rq\rq. In fact, this was discussed above,  see Eq.~(\ref{cs}). Another
central charge is related to the junction \lq\lq hub\rq\rq. 

In what follows we will assume the circumference of the world sheet cylinder to
be large. This is sufficient to ensure the applicability of the quasi-classical
approximation. 

In the quasi-classical limit $L\rightarrow \infty$ the central charge related
to the junction is subdominant. From Fig.~\ref{fig3I}, it is evident that at
$L\rightarrow \infty$ the minimal action is
\begin{equation}
\Delta A_{\rm min} = L \sigma,
\end{equation}
where $\sigma$ is the tension of the horizontal \lq\lq wall\rq\rq,
\begin{equation}
\sigma = 2 (I_{*r} - I_{*l}).
\end{equation}
The effect of the \lq\lq hub\rq\rq\ is subdominant, it is proportional to $L^0$,
and may be of the same order as the pre-exponential factors due to the 
zero modes.
Thus,
\begin{equation}
M-|Z| \propto e^{-4(I_{*r} - I_{*l}) L}.
\end{equation}
In the next section we will consider a concrete model, which seems to present
an instructive example. In this particular model we calculate for $\Delta
A_{\rm min}$ both  the linear term in $L$ and, for the sake of completeness,
the next to leading term associated with the $(1/2,1/2)$ central charge, the
\lq\lq hub\rq\rq.

The remainder of the paper presents an illustration and elaboration of the above
general results.

\section{An (instructive) example}

In this section we apply the previous considerations to a specific model which
was first introduced in~\cite{HLS}. We consider a generalized Wess-Zumino model
for which
\begin{equation}
{\mathcal K}(\Phi,\bar\Phi)=\Phi\bar\Phi, \qquad d
{\mathcal W}=\frac{4\pi}{2-\cos\Phi}d\Phi,
\label{oldsup}
\end{equation}
where $d{\mathcal W}/d\Phi$ is a single-valued function derived from the 
multi-valued superpotential
\begin{equation}
{\mathcal W}=\frac{8\pi}{\sqrt3}\arctan\left(\sqrt 3\tan\frac\Phi2\right).
\label{superpot}
\end{equation}
The model possesses only a run-away vacuum
$\vert{\mathrm{Im}}(\phi)\vert\to\infty$. It is stabilized by solitons,  which
at the classical level are solutions to the BPS equation given by
Eq.~(\ref{BPSeq}). 

\begin{figure}[!t]
\begin{center}
\includegraphics[width=8.0cm]{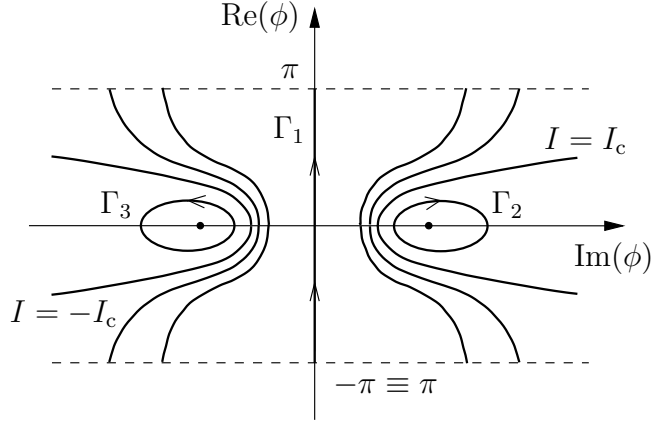}
\put(-7,21){Im$(\phi)$}
\put(-54,53){Re$(\phi)$}
\put(-47,38){$\Gamma_1$}
\put(-18,28){$\Gamma_2$}
\put(-70,28){$\Gamma_3$}
\put(-39,4){$-\pi\equiv\pi$}
\put(-46,46){$\pi$}
\put(-11,36.5){$I=I_{\rm c}$}
\put(-82,13.5){$I=-I_{\rm c}$}
\caption{The target space of the model. The three types of non-contractable cycles
are indicated by $\Gamma_1$ and $\Gamma_{2}$ and $\Gamma_{3}$.}
\label{fig1}
\end{center}
\end{figure}

The target space has the topology of a cylinder
($-\infty<{\mathrm{Im}}(\phi)<+\infty,\ -\pi\le{\mathrm{Re}}(\phi)\le\pi$), 
with the poles of the scalar potential,
\begin{equation}
(\phi_*)_{1,2}=\pm\imath\log\left(2+\sqrt3\right),
\end{equation}
removed. Each soliton solution belongs to one of  three homotopy classes,
$\Gamma_1$,  $\Gamma_2$ and  $\Gamma_3$. Solutions in  $\Gamma_1$ wind around
the target space cylinder, whereas solutions in $\Gamma_2$ and $\Gamma_3$ wind
around the points that are removed from the target space (see Fig.~\ref{fig1}).
Each solution in $\Gamma_1$ has a mirror image in the real axis of the complex
$\phi$ plane that belongs to the same class, except for a real soliton, which
is invariant under this transformation. Solutions in $\Gamma_2$ are mapped to 
solutions in $\Gamma_3$ and vice versa. The solutions in all homotopy classes
have equal period,
\begin{equation}
\Delta{\mathcal W}=\frac{8\pi^2}{\sqrt3}.
\end{equation}

\setlength{\unitlength}{.1cm}

As explained in Section \ref{sec:generalresults}, the constant of motion
$I={\rm Im} \left({\cal W} \right)$ (remember, $\delta=0$) may be used to mark
all solutions to the BPS equation. The BPS solitons in homotopy class
$\Gamma_2$ and $\Gamma_3$ are obtained for $I\in(-\infty,I_{*l})$ and 
$I\in(I_{*r},+\infty)$  respectively, with $I_{*l}=-I_c$ and $I_{*r}=+I_c$, and
\begin{equation}
I_c=\frac{8\pi}{\sqrt{3}} 
{\rm arctanh} \frac{1}{\sqrt{3}}. 
\end{equation}
The classically BPS saturated solitons in homotopy class $\Gamma_1$ are
obtained for $I\in(I_{*l},I_{*r})$. The  period function $\ell(I)$ was plotted in
Ref.~\cite{HLS}; this function is positive, it is symmetric under reflection in
$I=0$, and it monotonically  increases from $\ell(I) \downarrow 0$ at $I
\rightarrow -\infty$ to $\ell \rightarrow \infty$ at $I\uparrow I_{*l}$. 
Between
$I=I_{*l}$ and  $I=I_{*r}$, the period function reaches a minimum value $\ell=1$
for $I=0$, where the classically BPS saturated soliton is real.

For a given circumference $L$ of the compact dimension, the allowed solitons
with winding number $N$ are obtained from the equation $N \ell(I) = L$. The
energy of BPS saturated solitons is independent of $L$ and given by
\begin{equation}
M_{\mathrm{BPS}} = \left| Z \right| = \frac{16 N \pi^2}{\sqrt{3}}.
\end{equation}
In this paper we explicitly discuss solitons with winding number $N=1$, but the
results can be trivially extended to arbitrary values of $N$. In the present
model, there are two BPS saturated solitons if $L < 1$, one in homotopy class
$\Gamma_2$ and one in $\Gamma_3$. If $L>1$ there are four classically BPS
saturated solitons, two in homotopy class $\Gamma_1$ and one each in $\Gamma_2$
and $\Gamma_3$.

It turns out that for practical purposes, it is convenient to mark the class
$\Gamma_1$ solitons by the quantity $B$, the imaginary part of $\phi$ when the
real part of $\phi$ is equal to $\pi$. There is a one to one correspondence
between $B$ and $I$ that takes the form  \begin{equation} I = \frac{8
\pi}{\sqrt{3}} {\rm arctanh} \left(\frac{1}{\sqrt{3}} \tanh \frac{B}{2}
\right), \end{equation} so that for the critical solitons $B(\pm
I_c)\rightarrow \pm \infty$, and $B=0$ for $I=0$. 

\subsection{Non-BPS solitons for $L<1$.}

Even though there are no BPS solitons in homotopy class $\Gamma_1$ when $L<1$,
this does not preclude the existence of non-BPS solitons in the same homotopy
class.  The energy of such objects is above the BPS bound, but they are static
and topologically stable.

To address this issue we study static solutions to the second order equations
of motion. The kinetic energy is minimal for the shortest path in the complex
$\phi$ plane, which means straight lines connecting equivalent points. In
addition, the scalar potential has a saddle point at the origin in the target
space and has ridges originating from this saddle point on the real axis. This
means that there is a static real solution to the second order equation of
motion for any value of $L$. For $L=1$ such a solution saturates the BPS
bound.  For $L < 1$, the kinetic energy dominates and the total energy is
minimized by the straight line on the real axis in the complex $\phi$ plane.
Moreover, there are no BPS saturated solitons of the same homotopy class in
this regime; the real soliton is stable because there is nothing to  decay
into. For $L>1$,  the kinetic energy does not dominate any more; there are
other static solutions that are not straight lines in the complex $\phi$ plane
that actually have lower total energy, the BPS saturated complex solitons. In
this regime,  the real soliton is unstable. We will discuss it  in the next
section. 

\begin{figure}[!t]
\begin{center}
\includegraphics[width=7.0cm]{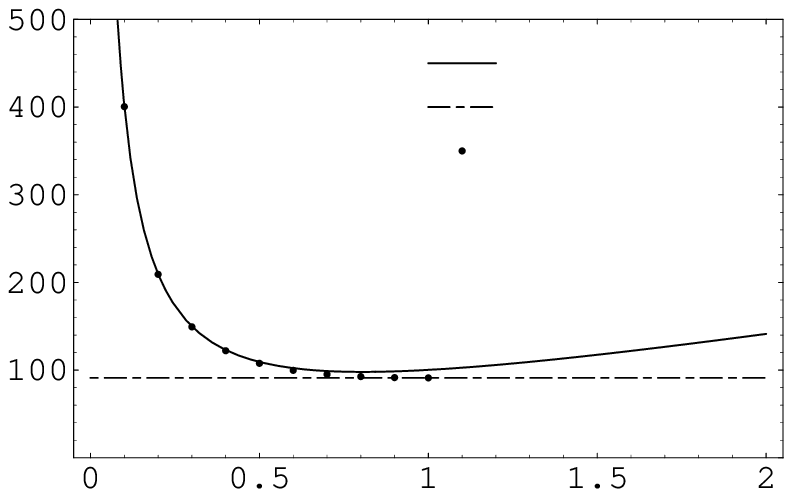}\hspace{12pt}
\includegraphics[width=7.0cm]{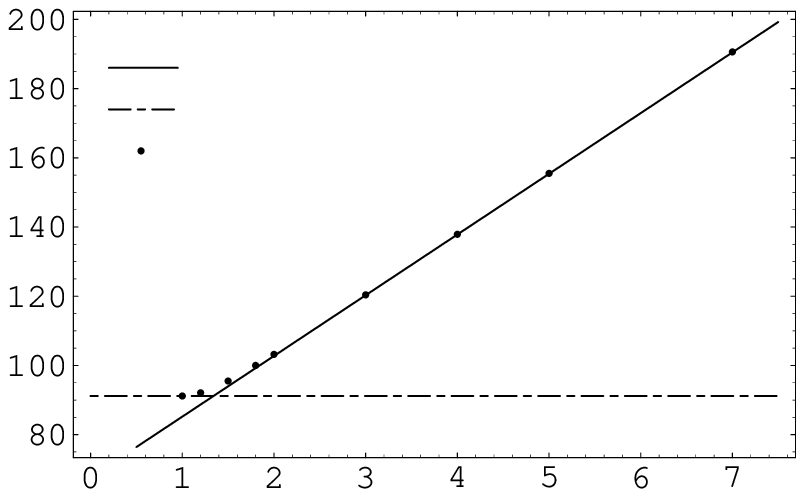}
\put(-200,58){\footnotesize{Numerical}}
\put(-200,66){\footnotesize{$M_{\textrm{\tiny{BPS}}}$}}
\put(-200,74){\footnotesize{$M_{\textrm{\tiny{sol}}}$ (approx.)}}
\put(-61,-7){$L$}
\put(-217,-7){$L$}
\put(-286,-7){\footnotesize{a)}}
\put(-138,-7){\footnotesize{b)}}
\put(-296,40){\footnotesize{\rotateleft{Energy}}}
\put(-147,40){\footnotesize{\rotateleft{Energy}}}
\put(-106,58){\footnotesize{Numerical}}
\put(-106,66){\footnotesize{$M_{\textrm{\tiny{BPS}}}$}}
\put(-106,73){\footnotesize{$M_{\textrm{\tiny{sphal}}}$ (approx.)}}
\caption{a) Energy of the real, non-BPS soliton as a function of the
circumference $L$ of the compact dimension; the BPS bound (dashed line), 
the analytical
approximation $M_{\rm sol}=U+K$ for small $L$ (solid line), 
and the actual numerically
calculated soliton mass (dots) are shown. b) Energy of the sphaleron, the real,
unstable soliton for $L>1$ as a function of the circumference $L$ of the
compact dimension; the BPS bound (dashed line), the linear analytical 
approximation for
$M_{\textrm{\tiny{sphal}}}$ for large $L$ (solid line), and the actual 
numerically calculated sphaleron mass (dots) are shown.}
\label{fig2}
\end{center}
\end{figure}

For $L \ll 1$, the real non-BPS soliton is approximately given by
\begin{equation}
\phi(x)=2 \pi \frac{x}{L},
\end{equation}
where $x$ ranges between $-L/2$ and $L/2$.
For this configuration, the kinetic energy is given by
\begin{equation}
K=\int_{-L/2}^{L/2}\!dx\,\left(\frac{2 \pi}{L}\right)^2  
=\frac{4\pi^2}L,
\end{equation}
while the potential energy is
\begin{equation}
U=\int_{-L/2}^{L/2}\!dx\,\frac{16\pi^2}{(2-\cos 2 \pi x/L)^2}= 
\frac{32 L \pi^2}{3 \sqrt{3}}.
\end{equation}
For $L\ll 1$, the energy of the soliton therefore approaches $M_{\rm sol}=U+K$.
In Fig.~\ref{fig2}a) we show the BPS bound and  the approximate soliton energy
for small $L$, together with the numerically calculated energy of the non-BPS
soliton.

\setlength{\unitlength}{.05cm}
  
\subsection{Sphaleron for $L>1$.}

The real, static solution of the second order equation of motion persists even
for $L>1$. In this regime, the soliton is unstable and we will refer to it as a
sphaleron, by analogy with the sphaleron in the Yang-Mills theory~\cite{SPHYM}.
As in the Yang-Mills theory, the sphaleron mass in our problem will give the
height of the barrier under which the solitons $(l)$ tunnels into $(r)$ 
and {\em
vice versa}.  Because the solution is real, the second order equation of motion
with vanishing time derivative can be integrated. In fact, the equation of
motion is identical to the equation describing the one-dimensional motion of a
particle moving in the  potential $-V(x)$. The implicit solution for $\phi$ is
\begin{equation}
x-x_0 =\int_{0}^{\phi} d\theta /\sqrt{V(\theta)+V_0},
\end{equation}
where $V_0$ is an integration constant (equivalent to the total energy of the
particle).  The solution $\phi(x)$ is periodic modulo $2\pi$ with wavelength
$\ell(V_0)$. The constant $V_0$  has to be adjusted so that the wavelength of
the solution (which corresponds to the ``time" it takes for the particle to
move around the circle) is equal to the circumference $L$ of the compact
dimension, that is $\ell(V_0)=L$. This equation has one solution for any
positive value of $L$. The method to  determine  $V_0$ is similar to the
procedure that was used to select $I$ in the case of the BPS solitons. The
energy of the real soliton/sphaleron is equal to 
\begin{equation}
M_{\rm sphal}= \int_{-\pi}^{\pi} d\phi\, 2 \sqrt{V(\phi)+V_0} - V_0 L.
\end{equation}
This energy can be explicitly determined in various limits. For $V_0 \gg 0$,
the real soliton of the previous section is obtained, with $\sqrt{V_0}=2\pi/L$
(this corresponds to a particle with so much kinetic energy that it hardly
notices the potential), whereas for  $V_0=0$ the solution is equal to the real
BPS saturated soliton with $L=1$ and energy equal to the BPS bound. When $V_0$
is close to minus the minimum value of the potential, $V_0 \approx -16\pi^2/9$,
then $L \gg 1$ and the energy increases linearly in $L$,
\begin{equation}
M_{\rm sphal}= \frac{ 8 \pi^2}{\sqrt{3}} + \frac{16 \pi}{3} \log\frac{\sqrt{3}+1}{\sqrt{3}-1} + 
\frac{16\pi^2 }{9} L + \cdots,
\label{sphalen}
\end{equation}
where the ellipsis indicate terms that vanish in the limit $L\rightarrow
\infty$. (This last situation corresponds to a particle with just barely enough
energy to reach the top of the hill). In Fig.~\ref{fig2}b) we show the  mass of
the BPS saturated soliton and  the approximate sphaleron energy for large $L$
in Eq.~(\ref{sphalen}),  together with the actual numerically calculated energy
of the sphaleron.

\subsection{The Tunneling Action.}

For any value $L>1$, there are four classical BPS saturated solitons, two in
class $\Gamma_1$ and one each in class $\Gamma_2$ and  $\Gamma_3$ .  The two
solitons in class $\Gamma_1$ are marked by  values of $B$ that have the same
magnitude but opposite sign. They are mapped onto each other in the complex
$\phi$ plane by reflection in the real axis. In order to distinguish these two
solitons, we will refer  to them as the $(l)$ soliton when $B$ is negative, and
the $(r)$ soliton when $B$ is positive. Tunneling mixes the  two class
$\Gamma_1$ solitons, and, as a consequence, their mass is lifted above the BPS
bound. The class $\Gamma_2$ and $\Gamma_3$ BPS solitons do not mix;  the
tunneling action is infinite since the cycle would have to be moved across the
poles (see Fig.~\ref{fig1}); but the $(l)$ and the $(r)$ solitons can be
deformed into each other without crossing a pole. The energy barrier that
separates them is therefore finite, see Fig. 5.  In addition, at $L\gg1$ the
barrier is high, and the quasi-classical approximation is applicable. According
to our previous considerations, the two shortened supermultiplets  pair up to
form a full representation, and the mass is lifted from the BPS bound.  

\subsubsection{BPS bound on the tunneling action}

Here we will derive the BPS bound on the tunneling action for the specific
model under consideration using the methodology outlined in Section
\ref{sec:generalresults}. The bound is saturated if the instanton
configuration that interpolates between the $(r)$ soliton at $T=-\infty$ and
the $(l)$ soliton at $T=\infty$ satisfies the two-dimensional BPS equation
given in Eq.~(\ref{BPSjunc}); in Section \ref{numsim} we will use numerical
methods to show that the instanton is indeed BPS saturated. 

In order to determine the surface terms in Eq.~(\ref{euclact}), the BPS bound on
the  Euclidean action can be written as
\begin{equation}
A_{{\rm BPS}}=\int dt\,dx\,\left[2\vec{\nabla}\cdot\vec{S}
-(\vec\nabla\wedge\vec a)_z \right],
\label{surint}
\end{equation}
where
\begin{equation}
\vec{S}= 
\left[ 
\begin{array}{c}
{\rm Re}\left({\mathcal{W}}\right) \\
{\rm Im}\left({\mathcal{W}}\right)
\end{array}
\right], \qquad
\vec{a}=-
\left[ 
\begin{array}{c}
{\rm Im}(\phi\partial_x\bar{\phi})\\
{\rm Im}(\phi\partial_t\bar{\phi})
\end{array}
\right],
\label{def}
\end{equation}
and we have used  $(\vec \nabla \wedge \vec a)_z$ as short-hand for 
$\partial_x a_t - \partial_t
a_x$.
Then application of Gauss' and Stoke's theorems converts the surface integral 
in 
Eq.~(\ref{surint}) into contour integrals over the boundaries of the surface, {\it
i.e.}
\begin{equation}
A_{{\rm BPS}} = 2 \oint \vec{S} \cdot d\vec{n}
- \oint \vec{a} \cdot d\vec{x}
\label{countint}
\end{equation}

As noted in Section~\ref{sec:generalresults}, this is the same equation
derived  in Ref.~\cite{GS}  for the BPS bound on the energy of domain wall
junctions. In the first integral in Eq.~(\ref{countint}), $d\vec{n}$ is an
infinitesimal vector perpendicular to the contour with length $\vert d\vec
x\vert$, pointing outwards from the enclosed area. In the second integral,
$d\vec{x}$ is an infinitesimal vector tangential to the contour, and the
contour must be followed counter-clockwise.

\begin{figure}[!t]
\begin{center}
\includegraphics[width=6.0cm]{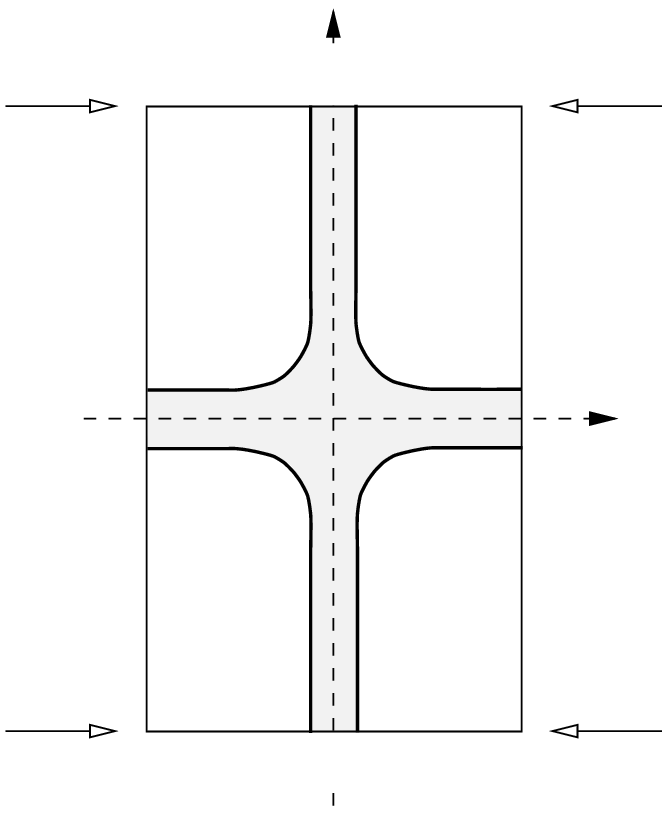}
\put(-25.5,9){$-T/2$}
\put(-12,31){$L/2$}
\put(-44,59.5){$T/2$}
\put(-58,37.5){$-L/2$}
\put(-72,3){\small{$\phi=-\pi-\imath B(L)$}}
\put(-15,3){\small{$\phi=\pi-\imath B(L)$}}
\put(-72,65.5){\small{$\phi=-\pi+\imath B(L)$}}
\put(-15,65.5){\small{$\phi=\pi+\imath B(L)$}}
\put(-6,36.5){$x$}
\put(-33,70.5){$t$}
\put(-40,3){$(l)$ Soliton}
\put(-39,65.5){$(r)$ Soliton}
\caption{Boundary conditions on the world sheet. The initial $(l)$
and final $(r)$ solitons
are located at the bottom and top sides of the rectangle at $t=-T/2$ and
$t=T/2$, with the center of the energy density positioned at $x=0$. Periodic (modulo
$2\pi$) boundary conditions apply to the left and right side of the rectangle. The
shaded area significantly contributes to the action.}
\label{fig3}
\end{center}
\end{figure}

The problem of calculating the Euclidean action is therefore equivalent to the
calculation of the energy of a domain wall junctions, with the solitons
corresponding to domain walls. The BPS bound on the action is completely
specified by the boundary conditions. We have to deal properly with the fact
that the $x$ direction in our model is compact. In Fig.~\ref{fig3} we show the
boundary conditions in the $x,t$ plane; at $t=-T/2$ the field $\phi(x,t)$ is
equal to the $(l)$ soliton, $\phi(x,-T/2)=\phi_l(x)$, where the $(l)$ soliton
is positioned so that it has its maximum energy density at $x=0$. Similarly, at
$t=+T/2$ the field $\phi(x,t)$ is equal to the $(r)$ soliton,
$\phi(x,+T/2)=\phi_r(x)$, where the $(r)$ soliton is also positioned so that it
reaches its maximum energy density at $x=0$. We always have in mind the limit
that $T$ is very large. For $x=-L/2$ and $x=L/2$, periodic (modulo $2\pi$)
boundary conditions have to be imposed, $\phi(-L/2,t)=\phi(L/2,t)-2\pi$, as the
space dimension is compact. 

\setlength{\unitlength}{.1cm}
 
The contour of the integrals in Eq.~(\ref{countint}) follows the edges of the
rectangle in Fig.~\ref{fig3}.  Special care must be taken with the integrals
over the vertical edges, at $x=\pm L/2$. Naively, one might think that these
contributions vanish, but because both the superpotential and the field are
multi-valued, these integrals do in fact contribute.

Before calculating the integrals in Eq.~(\ref{countint}), let us first determine
the dominant contribution from the vertical wall in the large $L$ limit. As
stated in Section~\ref{sec:generalresults}, this contribution is just $L$ times
the tension $\sigma$ of the horizontal ``wall''  in  Fig.~\ref{fig3}. If $L$ is
large, then the absolute value of $B$, the parameter that marks the $(r)$ and
$(l)$ solitons, becomes large. In fact, the absolute value of $B$ increases
logarithmically with $L$. Except for points in space near the center of the
soliton where the energy density is maximal, the soliton field is approximately
equal to $\pm \pi \pm \imath B$, where the second $\pm$ refers to the $(r)$ and
$(l)$ soliton, respectively. For $\delta=\pm \pi/2$, there is a solution to the
BPS equation that takes  the form $\phi=\pm \pi + \imath f(t)$, where $f(t)\to
\pm \infty$ for $t\to \pm \infty$. The tension of the horizontal ``wall'' 
$\sigma$ is  equal to the tension of this solution. The leading contribution to
$\Delta A$, linear in $L$, is therefore
\begin{equation}
\Delta A= \sigma L + \cdots =
\frac{32 \pi L}{\sqrt{3}}\,  {\rm arctanh} \left( \frac{1}{\sqrt{3}} \right) + \cdots, 
\label{domcon}
\end{equation}
where the ellipsis indicate subleading terms in $L$. We will now uncover these
subleading terms, which contribute to the pre-exponential factor in the
tunneling amplitude. We will first determine the contributions from the spokes
in Fig.~\ref{fig3} and then the hub.

\subsubsection*{Spoke contributions}

We first simultaneously calculate the contributions of the first integral in
Eq.~(\ref{countint})  over the left and right edges of the rectangle in
Fig.~\ref{fig3},
\begin{equation}
A_1 =  2 \int_{-T/2}^{T/2}\!\!\!\!dt\,\left[
\left.{\rm Re}\left({\mathcal W}\right)\right|_{x=L/2} - 
\left.{\rm Re}\left({\mathcal W}\right)\right|_{x=-L/2}\right].
\end{equation}
In order to calculate $A_1$, it is necessary to exploit some symmetries of the
soliton solutions. Apart from the translational invariance (modulo $2 \pi$)
\begin{equation}
\phi(x+L)= 2\pi + \phi(x),
\end{equation}
the soliton solutions have the following two symmetry properties
\begin{equation}
{\rm Re} \left[\phi(L/2+x)\right]=2\pi - {\rm Re}\left[\phi(L/2-x)\right], 
\label{specsym1}
\end{equation}
and
\begin{equation}
{\rm Im} \left[\phi(L/2+x)\right]={\rm Im} \left[\phi(L/2-x)\right].
\label{specsym2}
\end{equation}
If the symmetry in Eq.~(\ref{specsym1}) is preserved by the instanton
configuration, the interpolating field necessarily takes the form 
\begin{equation}
\begin{array}{l}
\phi(L/2,t)=\pi + \imath B(L,t) \\
\phi(-L/2,t)=-\pi + \imath B(L,t), 
\end{array}
\label{formphi}
\end{equation}
at the left and right edges of the rectangle in Fig.~\ref{fig3}, at $x=\pm
L/2$,  where $B(L,t)$ is a function interpolating between $-B(L)$ at $t=-T/2$
and $B(L)$ at $t=T/2$.  However, from the identity
\begin{equation}
{\mathcal W}(\pm \pi + \imath B) \equiv
\frac{8\pi}{\sqrt{3}}\left[ \pm \frac{\pi}{2} + \imath\,{\rm arctanh} \left(\frac{1}{\sqrt{3}}
\tanh \frac{B}{2} \right)\right],
\end{equation}
it is clear that the difference between the two integrands above does not
depend on the function $B(L,t)$ at all.  The contribution $A_1$ is therefore
given by
\begin{equation}
A_1 = \frac{16 \pi^2}{\sqrt{3}} T.
\end{equation}
This is just $T$ times the BPS bound on the soliton mass, or the Euclidean
action in the absence of the tunneling transition. 

Next, we simultaneously calculate the contributions of the first integral in
Eq.~(\ref{countint})  over the top and bottom edges of the rectangle in
Fig.~\ref{fig3},
\begin{equation}
A_2 =
2 \int_{-L/2}^{L/2}\!\!\!\!dx\,\left[
\left. {\rm Im}\left({\mathcal W} \right) \right|_{t=T/2}-
\left. {\rm Im}\left({\mathcal W} \right) \right|_{t=-T/2}\right].
\end{equation}
Here the integrands are just the constants of motion for the soliton 
solutions, so that $A_2 = 2 I_r - 2 I_l=4 I_r$, and $I_r$ and $I_l$ mark the
$(r)$ and the $(l)$ soliton, respectively.  Therefore, in terms of $B(L)$, the
value of $B$ that marks the $(r)$ soliton for a compact dimension with
circumference $L$, we obtain
\begin{equation}
A_2 = \frac{32 \pi}{\sqrt{3}}\, L\, 
{\rm arctanh} \left( \frac{1}{\sqrt{3}} \tanh \frac{B(L)}{2}\right).
\end{equation}
For large $L$, such that $B(L)\gg2$, this reduces to $A_2= \sigma L$, the
dominant contribution to the action that was already obtained in
Eq.~(\ref{domcon}).

\subsubsection*{Hub contributions}

We first simultaneously calculate the contribution of the second integral
in Eq.~(\ref{countint}) over the vertical edges in Fig.~\ref{fig3},
\begin{equation}
A_3=\int_{-T/2}^{T/2}\!\!\!\! dt\, \left[\left. {\rm Im}\left(
\phi \partial_t \bar{\phi} \right) \right|_{x=L/2}
-{\rm Im}\left.\left(\phi \partial_t \bar{\phi} \right) 
\right|_{x=-L/2}\right].
\end{equation}
Using the form of $\phi$ for $x=\pm L/2$ given in Eq.~(\ref{formphi}), this
contribution can be calculated and yields
\begin{equation}
A_3 = -2 \pi \int_{-T/2}^{T/2} dt\,  \partial_t B(L,t)=-4\pi B(L).
\end{equation}

\begin{figure}[!t]
\begin{center}
\includegraphics[width=6.0cm]{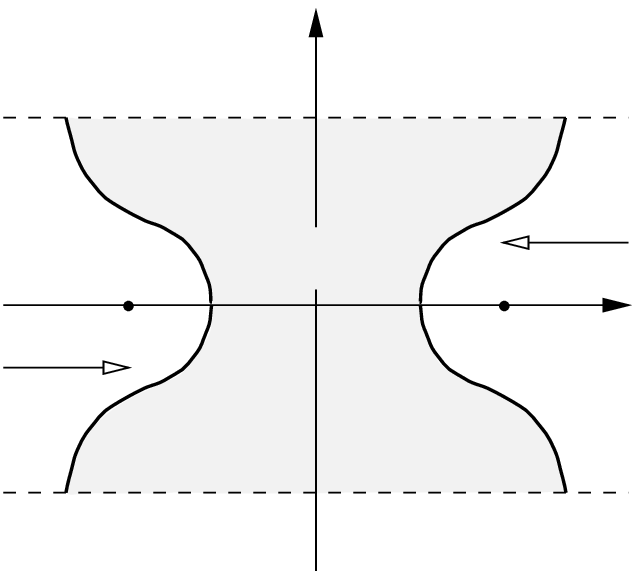}
\put(-34,45){$\pi$}
\put(-28,4){$-\pi$}
\put(-44,52){Re$(\phi)$}
\put(-6,21){Im$(\phi)$}
\put(-59,45){$-B(L)$}
\put(-11,45){$B(L)$}
\put(-7,28){$(r)$ Soliton}
\put(-72,16){$(l)$ Soliton}
\put(-34,29){$S(L)$}
\caption{The $(r)$ and $(l)$ solitons and the instanton in the complex $\phi$ plane
(compare Fig.(\ref{fig3}) for the same configuration on the world sheet).
The area $S(L)$ is indicated by the shaded
region.}
\label{fig4}
\end{center}
\end{figure}

Finally, we simultaneously calculate the contribution of the second integral
in Eq.~(\ref{countint}) over the horizontal edges in Fig.~\ref{fig3},
\begin{equation}
A_4 = 
\int_{-L/2}^{L/2}\!\!\!\! dx\,\left[ \left. {\rm Im} \left(
\phi \partial_x \bar{\phi} \right) \right|_{t=-T/2} -
\left. {\rm Im} \left(
\phi \partial_x \bar{\phi} \right) \right|_{t=T/2}\right].
\end{equation}
The $(r)$ and $(l)$ soliton are mapped onto each other by the transformation
\begin{equation}
{\rm Im}[\phi(x)] \rightarrow - {\rm Im}[\phi(x)].
\end{equation}
At the top and bottom edges of the rectangle in Fig.~\ref{fig3}, the field
$\phi$ therefore takes the form
\begin{equation}
\begin{array}{l}
\phi(x,T/2)=a(x)+\imath b(x), \nonumber \\
\phi(x,-T/2)=a(x)-\imath b(x),
\end{array}
\end{equation}
where $a(x)$ and $b(x)$ are real, and $a(\pm L/2)=\pm\pi$ and $b(\pm
L/2)=B(L)$. The integral can now be calculated and yields
\begin{equation}
A_4=-2 \int_{-L/2}^{L/2} dx\, \left( a\partial_xb - b\partial_x a \right)
= 4 \pi B(L) - 2 S(L),
\end{equation}
where 
\begin{equation}
S(L) = 2\int_{-L/2}^{L/2}dx\,b\partial_xa =
2\int_{x=-L/2}^{x=L/2}da\, b  =
2\int_{-\pi}^{\pi}da\, b,
\end{equation}
is the area in the complex $\phi$ plane  between the $(l)$ and $(r)$ soliton
configurations (see Fig.~\ref{fig4}).

The hub contribution to the action is thus seen to be equal to minus twice the
area traced by the instanton in the complex $\phi$ plane,  $A_{\rm
hub}=A_3+A_4=- 2 S(L)$. In Ref.\cite{CHT} it was
shown, within the context of domain wall junctions, that this result is valid
in general and goes beyond  the specific model we consider here.
We also note that $A_{\rm hub}$ is negative, in
agreement with the general considerations in  Ref.~\cite{SV}. The total
semi-classical Euclidan tunneling action is
\begin{eqnarray}
\Delta A_{\mathrm{BPS}} & = & A_2 + A_3 + A_4, \nonumber \\
 & = & \frac{32 \pi}{\sqrt{3}}\, L\, 
{\mathrm{arctanh}} \left( \frac{1}{\sqrt{3}} \tanh \frac{B(L)}{2}\right)
- 2 S(L).
\label{deltaA}
\end{eqnarray}
We have calculated each contribution in terms of the functions $B(L)$ and
$S(L)$. These functions depend only on the initial and the final soliton
configuration. In the large $L$ limit they take the form
\begin{equation}
S(L) =  2\pi\log L + \cdots, 
\end{equation}
and
\begin{equation}
B(L) = \log L + \cdots,
\label{Bscaling}
\end{equation}
where the ellipsis indicate terms that are finite or suppressed. We observe
that the hub contribution depends logarithmically on $L$, in contrast with the
situation for the domain wall junctions in models with infinite space
dimensions, for example those considered in \cite{BV,SV}, where the  hub
contribution to the energy of the junction is finite. The difference is that
for those models the energy density falls off exponentially fast away from the
domain walls, whereas in the present model the energy density only falls off
like the second power of the inverse distance to the wall. The circumference
$L$ of the compact dimension in a sense acts like an infrared  regulator. In
the large $L$ limit, the dominant and subdominant terms in tunneling action are
given by 
\begin{equation}
\Delta A \sim
\frac{32 \pi}{\sqrt{3}}\, L\, 
{\mathrm{arctanh}}  \frac1{\sqrt{3}} 
- 4\pi \log L + \cdots,
\end{equation}
where the subdominant term contributes to the pre-exponential factor in the
tunneling amplitude.

\subsubsection{BPS saturation of the instanton configuration}
\label{numsim}

The calculation in the previous section of the tunneling action hinges on the
question whether a BPS saturated instanton configuration exists. The explicit
form of such a configuration is not needed. In this section we use numerical
analysis to address the question if a BPS saturated instanton configuration 
exists in this specific model.  In order to numerically determine the instanton
configuration, we embedded the  model in $d=2+1$ dimensions. One of the  space
dimensions is compact and the other space dimension represents the original
Euclidean time. The new time is just an auxiliary construction.  An initial
configuration smoothly interpolates between the $(l)$ and $(r)$ solitons. The
second order equations of motions are then evaluated and at the same time the
system is cooled.  The instanton configuration then emerges as a domain wall
junction. 

We simulated the second order equations of motion on a lattice  using a forward
predicting algorithm similar to the one used in \cite{BV,SV}.  A complication
arises due to the fact that both  space  and the target space have the topology
of a cylinder, so that proper periodic boundary  conditions had to be
implemented.

\setlength{\unitlength}{.05cm}
\begin{figure}[!t]
\begin{center}
\includegraphics[width=8.0cm]{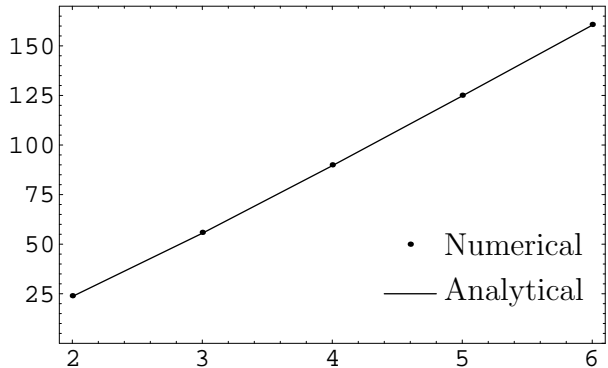}
\put(-43,32){Numerical}
\put(-43,20){Analytical}
\caption{Comparison between the BPS bound on the 
tunneling action from Eq.~(\ref{deltaA}) (solid line) and the actual semi-classical tunneling
action
from the numerical calculation (dots). It is clear that the actual instanton configuration
saturates the BPS bound.}
\label{fig5}
\end{center}
\end{figure}

The lattice spacing was chosen to be much smaller than the width of the
solitons. At the same time the size of the lattice was taken to be much larger
than the soliton width.  We performed the numerical calculation of the
tunneling action for values of $L$ ranging from $L=2$ up to $L=6$.  The
semi-classical tunneling action is only physically meaningful for larger values
of $L$. However, the computation time increases rapidly with $L$. As shown in
Fig.~\ref{fig5}, the  calculated value for the tunneling action with the
numerically determined instanton configuration saturates the BPS bound in
Eq.~(\ref{deltaA}). As there is no reason to think otherwise, we are confident
that this saturation also occurs for larger values of $L$.

\section{Quantum-mechanical description} 

Under certain conditions on the superpotential ${\cal W}$ (and the K\"{a}hler
potential $\cal{K}$, if it is non-trivial) it may be possible to develop a
{\em  quantum-mechanical} description of the tunneling of the distinct solitons
resulting in the loss of their BPS saturation.  Such an approach is valid
provided there is a certain direction in the functional space which is much
softer than all other ``perpendicular"  directions, and the tunneling occurs in
this ``soft" direction.  Then  all \lq\lq perpendicular\rq\rq\ degrees of
freedom (there are infinitely many of them) can be integrated out and what is
left is the quantum-mechanical motion of the center of mass plus the quantum
mechanical dynamics of this particular soft degree of freedom. The motion of
the center of mass is irrelevant, of course. It is always assumed that the
soliton is at rest. 

The example considered in Section 3 contains no adjustable parameters (except
$L$), so that one does not expect to find a specifically soft excitation mode
in the soliton background. Therefore, in this case the quantum-mechanical
description (with one bosonic variable) would describe the system only
qualitatively. It is not difficult, however, to modify the superpotential to
create a soft mode. To this end, let us consider the superpotential
\begin{equation}
d{\mathcal W}=\frac{4\pi}{2-\beta \cos\Phi}d\Phi\,,
\label{newsup}
\end{equation}
which introduces a parameter $\beta$ in the superpotential (\ref{oldsup}),
without changing anything else. At very small $\beta$ and large $L$,  
\begin{equation}
\log \frac{4}{\beta} \gg L \gg 1,
\end{equation} 
the $x$
dependence of the imaginary part of $\phi$  becomes weak, and the softest
mode corresponding to varying ${\rm Im} (\phi)$ is much softer
than all other modes. (We treat $L$ as a fixed parameter, while $\beta$ can
be made as small as desired.) Then we can define our
quantum  mechanical variable $B$ such that $B$ is the value of ${\rm Im}(\phi)$
(for definiteness, at  ${\rm Re}(\phi)=\pm \pi$). At the classical equilibrium
\begin{equation}
B_0 = \pm \log (L/\beta ) +{\rm const.}\, \qquad (L\gg 1)\,.
\label{dop}
\end{equation}
The positive value of $B_0$ corresponds to the $(r)$ soliton, the negative to
the $(l)$ soliton. Both are classically BPS saturated. To roll over from
positive  to negative $B$, the system has to tunnel under the barrier at
$B=0$. 

Before passing to the discussion of supersymmetric quantum mechanics for the
variable $B$ we have to present a few formula from Section 3 in a modified form
which includes the parameter $\beta$, see Eq.~(\ref{newsup}).

The superpotential with the additional parameter $\beta$ takes the form
\begin{equation}
{\cal W}= \frac{8 \pi}{\sqrt{4-\beta^2}} \arctan \left(
 \sqrt{\frac{2+\beta}{2-\beta}} \tan \frac{\phi}{2}
\right).
\end{equation}
The poles of the scalar potential are
located
on the imaginary axis at 
\begin{equation}
\phi = \pm \imath \log  \frac{2}{\beta}\left(1+\sqrt{1-\frac{\beta^2}{4}} \right),
\end{equation}
so at small $\beta$ they move further away from each other. The energy of the
classical BPS saturated soliton is
\begin{equation}
M_{\rm BPS} \equiv Z = \frac{16 \pi^2}{\sqrt{4-\beta^2}}\to 8\pi^2\,\,\,\mbox{at}\,\,\,
\beta\to 0.
\end{equation}
Finally, in the limit of large $L$ the sphaleron energy is equal to 
\begin{eqnarray}
M_{\rm sphal} &=& M_{\rm BPS}   -  \frac{32 \pi}{\sqrt{4-\beta^2}} 
\arctan \frac{\sqrt{2-\beta}}{\sqrt{\beta}} 
+ \frac{16 \pi}{2+\beta} 
\log \frac{\sqrt{2+\beta} + \sqrt{\beta}}{\sqrt{2+\beta} - \sqrt{\beta}} 
\nonumber \\
&&+ \frac{16 \pi^2}{(2+\beta)^2} L\nonumber\\[0.2cm]
&\to & M_{\rm BPS} +4\pi^2 L +{\rm const.}\,\,\,\mbox{at}\,\,\,
\beta\to 0.
\end{eqnarray}
We recall that the difference between the sphaleron energy and the classical
soliton mass measures the height of the  tunneling barrier.  The tunneling
action as a function of $\beta$ becomes
\begin{equation}
\Delta A = \frac{32 \pi L}{\sqrt{4-\beta^2}}
{\rm arctanh} \sqrt{\frac{2-\beta}{2+\beta}}\to 8\pi L\log\frac{4}{\beta}
\,\,\,\mbox{at}\,\,\,
\beta\to 0.
\end{equation}

Now we pass to the quantum-mechanical treatment. To begin, let us recall some
general aspects.  A  general algebraic consideration of the BPS soliton
dynamics in the extreme non-relativistic limit is carried out in \cite{RSVV}. 
We will adapt it here for our purposes.

The superalgebra we deal with has four supercharges and a central charge $Z$
(the latter is real in the problem at hand). In the extreme nonrelativistic
limit it can be represented as \cite{RSVV} (in the soliton rest-frame)
\begin{eqnarray}
Q_2^1 &=&\sqrt{2Z}\tau_1\otimes\sigma_3 +...\,,\qquad\quad  Q^2_1
=\sqrt{2Z}\tau_2\otimes\sigma_3+...\,,
\nonumber\\[0.3cm]
Q_1^1 &=& I\otimes \frac{1}{2\sqrt{m}}\left[ p \sigma_1 +W' (B)\sigma_2
\right]\,,\quad 
Q_2^2= I\otimes \frac{1}{2\sqrt{m}}\left[p \sigma_2 -W' (B)\sigma_1
\right]\,,
\label{qmalgebra}
\end{eqnarray}
where $$ p=-i d/dB$$ and $m$ is an effective inertia coefficient for the
variable $B$ (``mass"). The dots in the first line in Eq.~(\ref{qmalgebra})
stand for  higher order terms in the nonrelativistic expansion. Moreover,
$W(B)$ is the quantum-mechanical superpotential (not to be confused with the
field-theoretic  superpotential ${\cal W}$). The algebra (\ref{qmalgebra}) has,
as its subalgebra, Witten's quantum mechanics \cite{W}.

In principle, $W(B)$ could be calculated from the underlying field theory, in
the same way as it was done in a related problem in Ref. \cite{RSVV}. We will
not do this calculation in full, since our purpose here is mainly illustrative.
Instead, we will present a simplified  superpotential which properly conveys
qualitative features of the actual superpotential: (i) the existence of two
zeros of $W' (B)$ at $B=\pm |B_0|$ (double-well potential);
(ii) the existence of  the barrier at $B$
near zero,  with the maximum at zero, and with the height coinciding with 
$M_{\rm sphal} - Z= 4\pi^2 L$.

The corresponding superpotential and Hamiltonian have the form
\begin{eqnarray}
W & =& k\left(-\frac{B^3}3+ B_0^2 B\right),
\nonumber\\[0.3cm]
H &=& \frac{p^2}{2m} +\frac{1}{2m}(W')^2 + \frac{1}{{2m}}\sigma_3 W''\, ,
\label{qmspot}
\end{eqnarray}
where
\begin{eqnarray}
B_0 & = & \log L/\beta \nonumber \\
k & = & 6 \pi L \frac{\log 4/\beta}{\log^3 L/\beta}  \\
m & = & \frac92 L \frac{\log^2 4/\beta}{\log^2 L/\beta} \nonumber \label{par}
\end{eqnarray}
(Note that the overall two-by-two unit matrix following from the second line
in  Eq.~(\ref{qmalgebra}) is omitted in $H$. Its just replicates the states, in
two copies each. We will keep it in mind.)

$B_0^2$ is a function of $L$, which is logarithmically large and positive at
large $L$. As $L$ decreases, $B_0^2$ decreases too, and goes to zero at $L=1$.
(To see that this is the case, one should inspect the exact expression
for $B_0$ rather than the asymptotic form in Eq.(\ref{par}) valid
at large $L$.)
Below this point $B_0^2 <0$, and the Hamiltonian has no supersymmetric
solution  at the classical level. The Witten index of the system under
consideration is zero.

At positive $B_0^2$ there are two classical solutions of the equation
$W^{\prime} =0$, corresponding to two classical vacua. 
The tunneling between them was
thoroughly studied, see {\it e.g.} \cite{wqmt}. 
The one-instanton action is obviously
\begin{equation}
A_{\rm inst}=
\left[W(B=|B_0|) - W(B=-|B_0|)
\right]\equiv \Delta W = 
\frac{4k}{3}|B_0|^3.
\end{equation}
The instanton transition is accompanied by a fermion zero mode, which
suppresses all one-instanton amplitudes. The shift of the ground state from
zero is given by the instanton--anti-instanton transition and is proportional
to $\exp (-2A_{\rm inst})$ (for further details see \cite{wqmt}). This can also
clearly be seen from the Hamiltonian (\ref{qmspot}) which has a strictly
conserved  quantum number, $[\sigma_3 , H] = 0$. The one-instanton
transition would flip the spin. The ground state of the Hamiltonian 
(\ref{qmspot}) is doubly degenerate, with one spin-up and one spin-down state 
of energy $E\sim \exp (-2A_{\rm inst})$. If we include, in addition,  the
replication (which was mentioned above) due to the two-by-two unit matrix, we
get that the overall number of states is four, as it must be in any non-BPS
${\cal N}=2$ supermultiplet.

To conclude this section, we stress again that although the superpotential
specified in Eqs.(\ref{qmspot}) and (\ref{par}) correctly reproduces
gross features of the tunneling phenomenon, it is definitely not the
genuine superpotential that might arise in this problem should we
decide to actually calculate it. In particular, it vanishes at small
$\beta$  only logarithmically, while actually one could expect a power-type
suppression.

\section{Conclusions}
In a previous work, Ref.\cite{HLS}, it was observed that in 
certain $(1+1)$-dimensional models with 
${\cal N}=2$ supersymmetry and a compact space dimension, tunneling
between two distinct, classically BPS saturated solitons lifts their mass 
above the BPS bound. The two shortened  multiplets combine to 
form one regular ${\cal N}=2$
multiplet. One of the main new assertions in this paper is that the 
instanton that interpolates between the solitons in the
Euclidean time is a $1/4$ BPS saturated ``domain wall junction.'' The
semi-classical instanton action is determined by two central charges
in the supersymmetry algebra. The action can therefore be determined
solely from the solitons that are connected by the instanton; an explicit
instanton solution is not necessary.

Under certain conditions the description of the tunneling between the 
solitons in the field theory can be reduced to
a non-relativistic supersymmetric quantum mechanical model. 
This approach is valid if the direction in the functional space
in which the tunneling takes place is much softer than all other
directions (except for the zero modes), yet the tunneling action 
is large. The internal coordinate
along this direction then forms the one remaining degree of freedom after
all hard degrees of freedom have been ``integrated out.''

We illustrated the above observations explicitly in a specific model. The
semi-classical tunneling action was determined in the field theory. We used
numerical methods to verify that the instanton indeed corresponds to
the $1/4$ BPS saturated  domain wall junction.
We also determined a range of parameters for which the tunneling can be
described by a quantum mechanical model. Although possible in 
principle, we did not determine the actual superpotential of the quantum 
mechanical model in this range. Instead, we determined the parameters
of a toy superpotential that is similar to the actual
superpotential in all important features.

\section*{Acknowledgements}

M.S. and T.t.V. would like to thank A.~Losev for interesting discussions, and
D.B. would like to acknowledge useful discussions with V.~Vento.
This work was supported in part by the Department  of Energy under Grant No.
DE-FG02-94ER40823 and by National Science
Foundation under Grant No. PHY94-07194,  and by Ministerio de Educaci\'on y
Cultura under Grant No. DGICYT-PB97-1227.



\begin{thebibliography}{99}

\bibitem{DS}	G.~Dvali and M.~Shifman, Nucl.~Phys.~{\bf B 504}, 127 (1997),
		hep-th/9611213.

\bibitem{ADD}	N.~Arkani-Hamed, S.~Dimopoulos, G.~Dvali, Phys.~Lett.~{\bf B
		429}, 263 (1998), \mbox{hep-ph/9803315}; N.~Arkani-Hamed,
		S.~Dimopoulos, G.~Dvali, Phys.~Rev.~{\bf D 59}, 086004 (1999),
		\mbox{hep-ph/9807344}. 

\bibitem{MV}    Y.~Meurice and B.~Veneziano, Phys.~Lett.~{\bf B141}, 69 (1984).

\bibitem{ADS}   I.~Affleck, M.~Dine and N.~Seiberg, Nucl.~Phys.~{\bf B241},
                493 (1984); Phys.~Rev.~Lett. {\bf 52}, 1677 (1984).

\bibitem{HLS} 	X.~Hou, A.~Losev and M.~Shifman, Phys.~Rev.~{\bf D 61}, 
		085005 (2000), \mbox{hep-th/9910071}. 

\bibitem{AGIT}  J.~A.~de~Azc{\'a}rraga, J.~P.~Gauntlett, J.~M.~Izquierdo and
                P.~K.~Townsend, Phys.~Rev.~Lett.~{\bf 63}, 2443 (1989).

\bibitem{AT}	E.~R.~C.~Abraham and P.~K.~Townsend, Nucl.~Phys.~{\bf 351}, 313
		(1991); 
		M.~Cvetic, F.~Quevedo and S.~J.~Rey, Phys.~Rev.~Lett.~{\bf 67},
		1836 (1991).	

\bibitem{GT} 	G.~W.~Gibbons and P.~K.~Townsend, Phys.~Rev.~Lett.~{\bf 83},
		1727 (1999), \mbox{hep-th/9905196}.

\bibitem{CHT}	S. M. Carroll, S. Hellerman and M. Trodden, Phys.~Rev.~{\bf D
		61}, 065001 (2000), \mbox{hep-th/9905217}.

\bibitem{GS} 	A.~Gorsky and M.~Shifman, Phys.~Rev.~{\bf D 61}, 085001 (2000),
		\mbox{hep-th/9909015}.

\bibitem{BV} 	D.~Binosi and T.~ter~Veldhuis, Phys.~Lett.~{\bf B 476}, 124
		(2000), \mbox{hep-th/9912081}.

\bibitem{SV} 	M. Shifman and T. ter Veldhuis, {\it Calculating
                the tension of domain wall junctions and vortices in
                generalized Wess-Zumino models}, \mbox{hep-th/9912162}. 

\bibitem{CVIF}  S.~Cecotti, P.~Fendley, K.~Intriligator, C.~Vafa,
		Nucl.~Phys.~{\bf B 386}, 405 (1992), \mbox{hep-th/9204102}.

\bibitem{CS}	B. Chibisov and M. Shifman, Phys. Rev. {\bf D56}, 7990
                (1997), \mbox{hep-th/9706141}; Erratum-ibid. {\bf D58} 109901 (1998).

\bibitem{SPHYM}
R.~Dashen, B.~Hasslacher and A.~Neveu,
Phys.\ Rev.\  {\bf D10}, 4138 (1974);
F.~R.~Klinkhamer and N.~S.~Manton,
Phys.\ Rev.\  {\bf D30}, 2212 (1984).
For a review see V. Novikov {\em et al.},
{\em ABC of Instantons}, in M. Shifman, {\em ITEP
Lectures on Particle Physics and Field Theory}
(World Scientific, Singapore 1999), Vol. 1, 
p. 258. 

\bibitem{RSVV}  A.~Ritz, M.~Shifman,~ A.~Vainshtein and M.~Voloshin, to be published.

\bibitem{W}	E.~Witten, Nucl.~Phys.~{\bf B 188}, 513 (1981).

\bibitem{wqmt}
P.~Salomonson and J.~W.~van Holten,
Nucl.\ Phys.\  {\bf B196}, 509 (1982);\\
M.~Claudson and M.~B.~Halpern,
Nucl.\ Phys.\  {\bf B250}, 689 (1985).

\end{thebibliography}
\end{document}